\documentclass{jkas}

\def\year{2014} 
\def\month{October} 
\def\volume{47} 
\def\issue{0} 
\def\beginpage{1} 
\def\endpage{12} 
\def\received{---} 
\def\accepted{---} 
\date{Received \received; accepted \accepted}

\usepackage{flushend} 

\newcommand{\OIII}{\mbox{O\,\textsc{iii}}}

\newcommand{\OI}{\mbox{O\,\textsc{i}}}
\newcommand{\NII}{\mbox{N\,\textsc{ii}}}

\newcommand{\SII}{\mbox{S\,\textsc{ii}}}

\newcommand{\kms}{km s$^{-1}$}
\newcommand{\ergs}{erg~s$^{-1}$}
\newcommand{\Ha}{H$\alpha$}    
    
\newcommand{\Hb}{H$\beta$}    
\newcommand{\mbh}{M$_{\rm BH}$}

\def\mnras{MNRAS}
\def\apj{ApJ}
\def\apjs{ApJS}
\def\aj{AJ}
\def\pasp{PASP}

\def\etal{et al.}

\title{
Misclassified Type 1 AGNs in the Local Universe
}

\author[1]{Jong-Hak~Woo}
\author[1,2]{Ji-Gang~Kim}
\author[3]{Daeseong~Park}
\author[4]{Hyun-Jin~Bae}
\author[2]{Jae-Hyuk~Kim}
\author[2]{Seung-Eon~Lee}
\author[5]{Sang~Chul~Kim}
\author[2]{Hong-Jin~Kwon}

\affil[1]{Astronomy Program, Department of Physics and Astronomy, Seoul National University, Seoul 151-742, Korea; \email{woo@astro.snu.ac.kr}}
\affil[2]{R\&E Program, Gyeonggi Science High School for the Gifted, 68-23 Songjuk, Jangan, Suwon 440-800, Korea}
\affil[3]{Department of Physics and Astronomy, University of California Irvine,
CA, USA}
\affil[4]{Department of Astronomy and Center for Galaxy Evolution Research, Yonsei University, Seoul 120-749, Korea}
\affil[5]{Korea Astronomy and Space Science Institute, Daejeon 305-348, Korea}

\def\corrauthor{
J.-H. Woo
}

\def\runningauthor{
Woo et al.
}

\def\runningtitle{
Misclassified Type 1 AGNs
}

\def\keywords{
galaxies: active --- galaxies: nuclei --- galaxies: Seyfert: H$\alpha$ emission line
}

\def\abstracttext{
We search for misclassified type 1 AGNs among type 2 AGNs identified with emission line flux ratios, and investigate the properties of the sample. Using 4\,113 local type 2 AGNs at $0.02<z<0.05$ selected from Sloan Digital Sky Survey Data Release 7, we detected a broad component of the \Ha\ line with a Full-Width at Half-Maximum (FWHM) ranging from 1\,700 to 19\,090 \kms\ for 142 objects, based on the spectral decomposition and visual inspection. The fraction of the misclassified type 1 AGNs among type 2 AGN sample is $\sim$3.5\%, implying that a large number of missing type 1 AGN population may exist. The misclassified type 1 AGNs have relatively low luminosity with a mean broad \Ha\ luminosity, log L$_{H\alpha} = 40.50\pm0.35$ \ergs, while black hole mass of the sample is comparable to that of the local black hole population, with a mean black hole mass, log M$_{\rm BH} = 6.94\pm0.51$ M$_{\odot}$. The mean Eddington ratio of the sample is log L$_{\rm bol}$/L$_{\rm Edd}$ = $-2.00\pm0.40$, indicating that black hole activity is relatively weak, hence, AGN continuum is too weak to change the host galaxy color. We find that the \OIII\ lines show significant velocity offsets, presumably due to outflows in the narrow-line region, while the velocity offset of the narrow component of the \Ha\ line is not prominent, consistent with the ionized gas kinematics of general type 1 AGN population. 
}

\begin{document}
\jkashead 


\section{Introduction}

Large-area surveys performed in various wavelengths, i.e., X-ray, optical, near-infrared, 
and mid-infrared, provide a large sample of active galactic nuclei (AGNs), enabling  various statistical studies of actively mass accreting supermassive black holes. In particular, by providing the rest-frame optical and UV spectroscopic properties of more than 100\,000 AGNs over a large redshift range, the Sloan Digital Sky Survey (SDSS) has dramatically changed our understanding of AGN population
\citep[e.g.,][]{ab09}, including the local black hole activity \citep[see][]{Heckman14},
AGN luminosity and Eddington ratio functions from low- to high-redshift \citep[e.g.,][]
{Kelly13}, and the connection of black hole activity to star formation
\citep[e.g.,][]{Netzer09}.

However, all surveys have their selection functions, hence, it is challenging to avoid selection biases 
in providing a complete sample of AGNs. In the case of the SDSS, the color-color diagram based
on the imaging survey has been used for selecting potential AGN candidates as 
spectroscopic follow-up targets. 
Thus, obscured AGNs, i.e., red AGNs \citep[e.g.,][]{Glikman07,Glikman13}
and X-ray-bright-optically-normal AGNs \citep[e.g.,][]{Hornschemeier05}, 
can be easily missed from the survey. 

In the optical spectroscopic studies, type 1 and type 2 AGNs are often classified based on 
the presence or absence of broad-emission lines, which are usually defined  
having a FWHM larger than 1000 \kms, as initially recognized among nearby 
Seyfert galaxies \citep[e.g.,][]{Seyfert43}. However, if the broad component 
of the Balmer lines is relatively weak and/or the narrow
component is dominant, then it is likely that 
type 1 AGNs can be misclassified as type 2 AGNs. In addition, if the AGN continuum is relatively weak
compared to the stellar continuum, then these objects are likely to be classified as galaxies rather than AGNs in the color-color diagram. 

A number of statistical studies have been performed using the SDSS galaxy catalogues,
e.g., the MPA-JHU value-added catalog,\footnote{http://www.mpa-garching.mpg.de/SDSS/} and the KIAS value-added galaxy catalog \citep{ch10}, which contain
the flux-limited local galaxy sample out to $z\sim0.2$, to investigate  
the properties of type 2 AGN populations identified through the emission line flux ratios 
\citep{bal81,kew06}. It has been noticed that
some type 2 AGNs show a relatively broad component in the \Ha\ line, while 
a broad component is often missing in the \Hb\ line, presumably due to 
its weak flux compared to the stellar continuum, suggesting that at least
some fraction of type 2 AGNs identified through their emission line ratios could 
be genuine type 1 AGNs. A systematic search for these misclassified type 1 AGNs
is yet to be performed \citep[see, e.g.,][]{oh11,bw14}. 
These misclassified AGNs are interesting targets for further study since
they are likely to be low luminosity AGNs since their AGN continuum is relatively weak. 
At the same time, their host galaxy properties can be easily studied while the 
mass of the central black hole can be estimated from the broad component of \Ha\ 
using various single-epoch mass estimators \citep[e.g.,][]{wu02,par12,ben13}. 

In this paper, we search for misclassified type 1 AGNs using a large sample of local 
type 2 AGNs selected from SDSS DR7, by carefully examining the presence of the broad 
component of the \Ha\ line. Based on the newly found type 1 AGNs, we investigate 
the properties of black hole activity and the kinematics of the ionized gas.
Sample selection and the procedure for identifying type 1 AGNs are described in Sections 2 and 3, 
respectively. 
In Section 4, we present the properties of the newly found 142 type 1 AGNs and 
their gas properties.
Discussion and conclusions  are presented in Section 5. Throughout the paper, we used
the cosmological constants of $H_{0}=70$\,km\,s$^{-1}$\,Mpc$^{-1}$, $\Omega_{m}=0.3$, and $\Omega_{\Lambda}=0.7$.

\section{Sample Selection}

To select type 2 AGNs in the local universe, we utilized the MPA-JHU 
catalog, 
which contains 927\,552 galaxies from the SDSS DR 7 and their derived
properties.
We selected low-redshift galaxies (i.e., $0.02 < z < 0.05$) and 
excluded galaxies with low stellar velocity dispersion (i.e., below the
SDSS instrumental resolution, $\sigma$ $<$ 70 \kms) in order to utilize
the available stellar velocity dispersion measurements 
in comparing with the gas kinematics.  
Among these local galaxies, we selected emission-line objects
with signal-to-noise ratio S/N $\geq$ 3 for \Ha, \Hb, \OI, and \OIII, which were used 
to classify AGNs in the emission line flux ratio diagram \citep{bal81}. 
By using the demarcation line for AGNs and star-forming galaxies, 
0.73/[log \OI/\Ha\ +0.59 ] +1.33 $<$ log \OIII/\Hb\ or 
log \OI/\Ha\ $> -0.59$ \citep{kew06}, we selected 4\,113 objects as the type 2 AGN sample.

\section{Analysis}

\subsection{Spectral Decomposition}

\begin{figure}[!t]
\centering
\includegraphics[width=82mm]{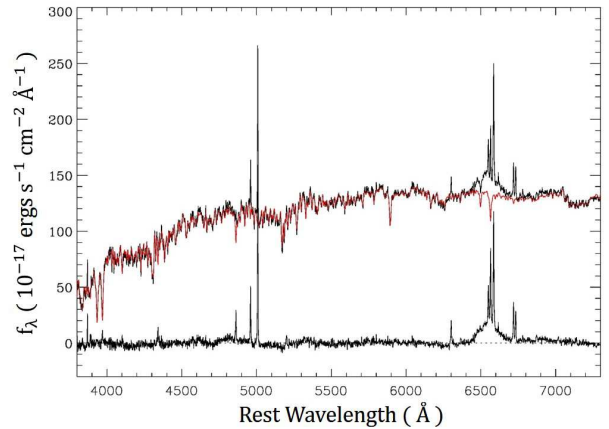}
\caption{Example of the spectral decomposition for SDSS J140543.16+251352.9.
The SDSS spectrum (black line) is modeled with a stellar population model (red line).
The residual spectrum at the bottom represents a pure AGN emission line spectrum.  } 
\label{fig1}
\end{figure}

In studying type 2 AGNs, it is of importance to decompose AGN emission lines
from the host galaxy stellar continuum in order to properly measure the flux and
width of each AGN emission line. For relatively low luminosity AGNs,
the line strength of stellar absorption lines is significantly large, hence 
the precise measurements of AGN emission line flux requires 
spectral decomposition of the stellar component, for example, 
using stellar population models or stellar spectral templates \citep[e.g.,][]{par12}. 
In addition, for measuring the kinematics of the ionized gas, it is necessary to measure the systemic velocity of the target galaxy based on the stellar absorption lines. 
In this study, we used the penalized pixel-fitting (pPXF) method \citep{cap04} 
to obtain the best fit stellar continuum model, which is based on the 235 
simple stellar population models 
(i.e., 5 different metallicities $\times$ 47 different ages) from the MILES library.
First, we de-redshifted all spectra using the redshift value from SDSS. After
fitting with the stellar continuum model, we refined the redshift (hence, the 
systemic velocity) of each object before the emission line fitting procedure. 
Then, the best-fit continuum model was subtracted from the SDSS spectra, 
leaving the pure AGN emission-line spectra. This approach is similar to 
a number of previous studies on the AGN emission and stellar absorption lines
\citep[e.g.,][]{woo10, par12, woo13}. 
We present an example of the spectral decomposition in Figure \ref{fig1}.

\begin{figure*}[t!]
\centering
\includegraphics[width=130mm]{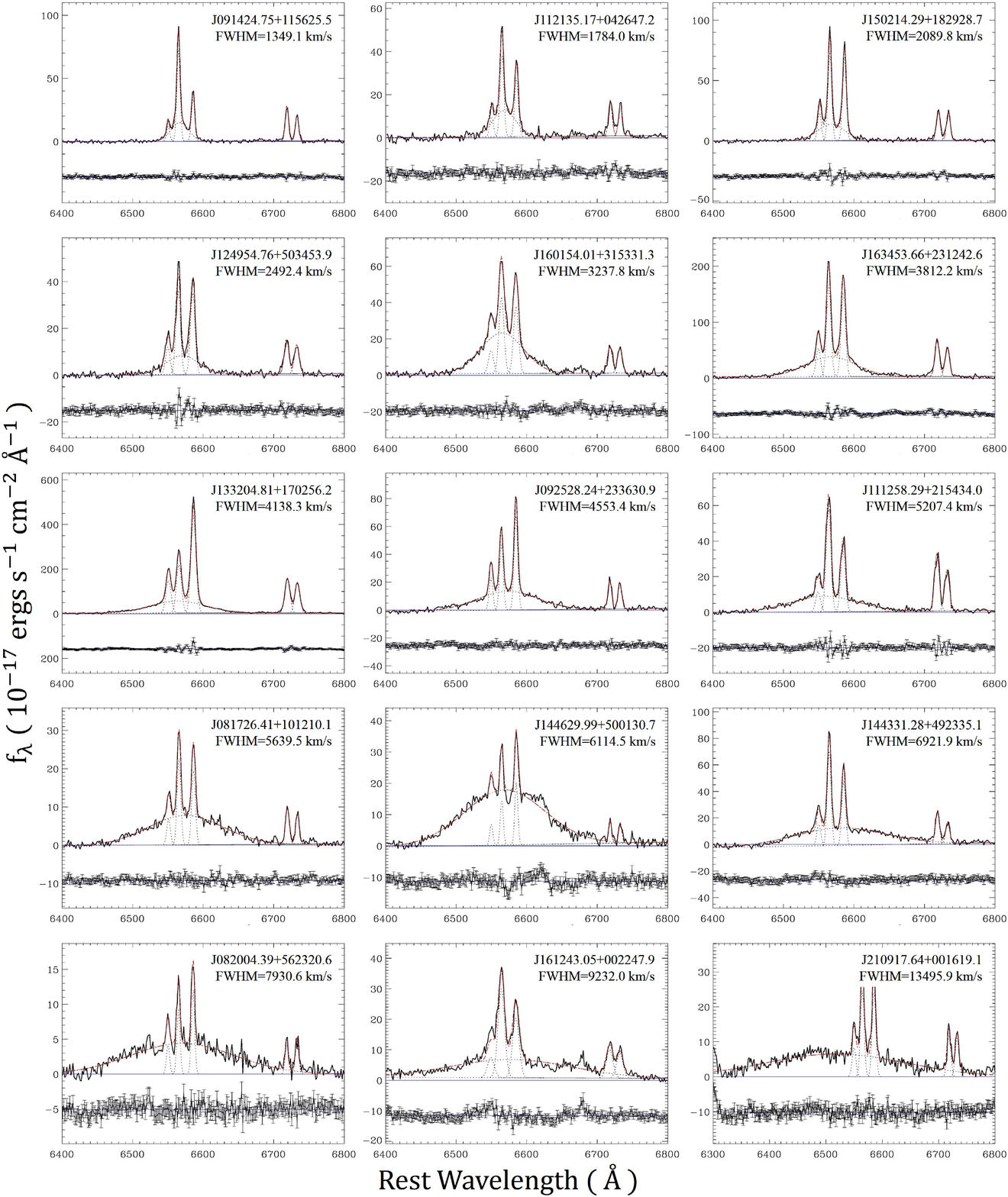}
\caption{Examples of the best-fit emission line models in the \Ha\ region 
for 15 objects, are presented in order of increasing FWHM of the broad 
\Ha\ component. 
Continuum-subtracted spectra (black lines) are plotted together with
the best-fit models (red lines), which are composed of 
narrow lines (\NII, \Ha, and \SII) and a broad \Ha\ component. 
At the bottom of each panel the residual is presented (black lines). 
The object name and the FWHM of the broad \Ha\ component are
given in each panel. 
}
\label{fig2}
\end{figure*}

\subsection{Emission Line Fitting}

We examined the SDSS spectra and continuum-subtracted spectra of 
all 4,113 objects in the type 2 AGN sample, in order to determine whether a broad 
component is present in the \Ha\ line. In this process,
we identified a sample of potential type 1 AGN candidates showing the broad \Ha\
component, based on the visual inspection.
Then, we carefully examined the \Ha\ region in the rest-frame 6300--6900\AA, 
by decomposing the broad and narrow components of \Ha, and \NII\ lines simultaneously.
If the stellar continuum was poorly subtracted from the SDSS spectra due to the low S/N 
ratio, a straight line was adopted to represent the continuum. 

Using the MPFIT routine \citep{mar09}, we modeled each narrow emission line in the \Ha\ 
region, namely, \Ha\ $\lambda\lambda6563$, \NII\ $\lambda\lambda6548$, $6583$ doublet, 
and \SII\ $\lambda\lambda6716$, $6731$ doublet with a single Gaussian component.
In addition, we added a Gaussian component with a FWHM $>$ 1000 \kms, to represent
the broad component of \Ha. 
The centers of each Gaussian components for narrow emission lines were fixed 
relative to each other 
at their laboratory separations, while the center of the broad \Ha\ component 
was set to vary in the fitting process. 
In the case of the line width, we used the same line dispersion value for all 
narrow emission lines, but for the broad \Ha\ component, we used a free parameter. 
The line flux ratio of [\NII] $\lambda6583$ and [\NII] $\lambda6548$ is fixed 
at the theoretical value of 2.96. 

\begin{figure*}[t!]
\centering
\includegraphics[width=160mm]{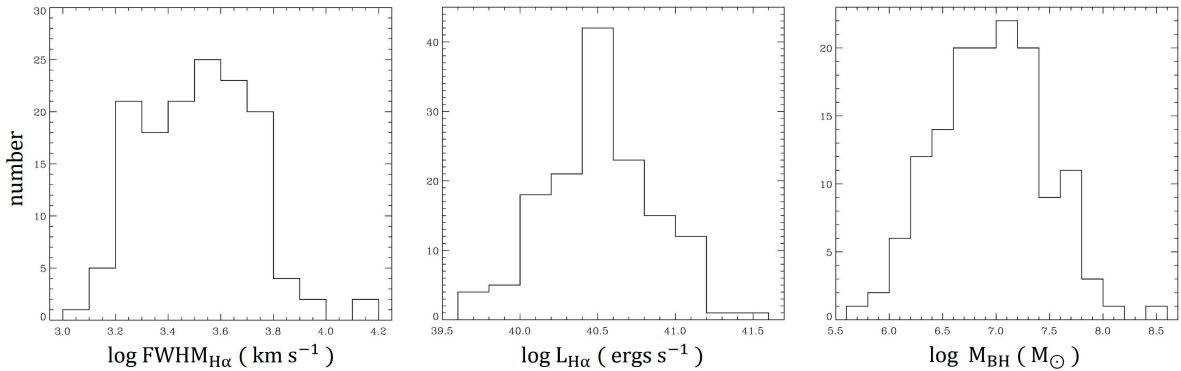}
\caption{Distribution of the measured FWHM$_{H\alpha}$ (left), log L$_{H\alpha}$ (center), log M$_{\rm BH}$ (right) for 142 misclassified type 1 AGNs.
}
\label{fig3}
\end{figure*}

Note that a broad residual from the incorrect subtraction of stellar continuum 
or the combination of the narrow \Ha\ and \NII\ line wings could be misinterpreted 
as a broad-line component. Thus, a careful examination is required
to avoid false detection of a broad \Ha\ component. 
For each object, we carefully examined and compared the raw spectrum
and the best-fit model,  and conservatively classified the target 
as type 1 AGNs only if the broad \Ha\ component is clearly needed for the fit. 
In particular, when the broad \Ha\ is relatively narrow (i.e., FWHM $<$ $\sim$2000 \kms), 
hence we do not see the wing of the broad \Ha\ blueward and redward of \NII\
in the continuum-subtracted spectra, we removed the target from the list of
the type 1 AGN candidates although the best-fit model includes a broad \Ha\ component. 
Thus, in our conservative classification it is possible that we may miss
relatively narrow broad-line objects (e.g., narrow-line Seyfert 1 galaxies).
In this process, we identified 142 AGNs, which clearly showed a broad \Ha\ component.
Figure \ref{fig2} presents examples of the emission line fit in the \Ha\ region for
15 AGNs in increasing magnitude of the FWHM of the broad \Ha\ component.

We measured the line flux and line dispersion of each narrow emission line and 
the broad \Ha\ line from the best-fit model. 
We used the luminosity distance using the redshift information in the header
of the fits file or the measured redshift from the stellar absorption lines.
These two redshifts showed negligible difference in luminosity distance.
Figure \ref{fig3} presents the distributions of the widths and luminosities of the broad
\Ha\ component (see Section 4.1 for details).

We also fit the \OIII\ line at 5007\AA, to measure the velocity center and the velocity
dispersion. Since the majority of the AGNs in the sample show a broad wing component
in the \OIII\ line profile, we used a double Gaussian model to fit the line profile. 
Then, we measured the flux centroid velocity and velocity dispersion of the best-fit
double-Gaussian model.

\subsection{Black Hole Mass and Eddington ratio}

Dynamical black hole mass estimation based on the spatially resolved kinematics is limited to nearby galaxies due to the limited spatial resolution \citep{kh13}.
For broad-line AGNs, black hole mass (\mbh) can be measured with the reverberation 
mapping method based on the virial assumption of the gas in the broad-line region (BLR) 
\citep{bla82}. Under the virial relation, black hole mass is expressed as M$_{\rm BH} = f~V^{2} R_{\rm BLR}/G$, where $V$ is the characteristic velocity scale of the broad-line gas, typically measured from the line dispersion of the Balmer lines, $f$ is the virial coefficient, which depends on the morphology, orientation, and the kinematics of the BLR, R$_{\rm BLR}$ is the size of the BLR measured from the reverberation mapping campaign, and $G$ is the gravitational constant.  

Adopting the BLR size -- luminosity relation from the recent calibration by \citet{ben13},
\begin{equation}\label{eq1}
R_{BLR}= 10^{1.527} \left[\frac{\lambda L_{5100}}{10^{44}\ {\rm erg\,s^{-1}}}\right]^{0.533} \rm light\ days
\end{equation}
the virial relation can be written as follows:
\begin{eqnarray}\label{eq2}
M_{BH} & = & f \times 10^{6.817} \bigg[\frac{V}{10^{3}\ {\rm km\,s^{-1}}}\bigg]^{2} \nonumber \\
 & ~ & \times\bigg[\frac{\lambda L_{5100}}{10^{44}\ {\rm erg\,s^{-1}}}\bigg]^{0.533} M_{\odot} ~ .
\end{eqnarray}
For the misclassified AGNs, we used the line width and luminosity of the broad \Ha\
for estimating black hole masses.
Adopting the calibration between the widths of \Hb\ and \Ha\ (i.e., FWHM$_{H\beta}$--FWHM$_{H\alpha}$ relation), and the relation between the AGN continuum luminosity 
and the \Ha\ line luminosity (i.e., $L_{5100}-L_{H\alpha}$ relation) from \citet{gre05},
\begin{equation}\label{eq3}
{\rm FWHM}_{H\beta}=1.07 \times 10^{3} \left[\frac{{\rm FWHM}_{H\alpha}}{10^{3}\ {\rm km\,s^{-1}}}\right]^{1.03} {\rm km\,s^{-1}} ~ ,
\end{equation}
we can derive M$_{\rm BH}$ from our measurements based on the spectroscopic decomposition
in the H$\alpha$ region. 
With the line dispersion ($\sigma_{H\alpha}$) and luminosity (L$_{H\alpha}$)
of the broad component of \Ha, the black hole mass can be estimated as follows:
\begin{eqnarray}\label{eq5}
M_{BH} & = & f \times 10^{6.565} \bigg[\frac{\sigma_{H\alpha}}{10^{3}\ {\rm km\,s^{-1}}}\bigg]^{2.06} \nonumber \\ 
 & ~ & \times\bigg[\frac{L_{H\alpha}}{10^{42}\ {\rm erg\,s^{-1}}}\bigg]^{0.46} M_{\odot} ~ .
\end{eqnarray}
For the virial factor we adopted $f=5.9^{+2.1}_{-1.5}$, which is based on 
the recent calibration using the combined sample of the quiescent galaxies and reverberation-mapped AGNs \citep{woo13}.
Note that the systematic uncertainty of the virial factor is 0.31 dex \citep[see][]{woo10}
and that black hole mass can easily vary by a factor of 2--3, depending on the virial 
factor calibration \citep[see][]{par12}. Thus, the quoted black hole mass should be
treated with caution. In Table~1, we provide the virial product, 
instead of the black hole mass, which can be determined by multiplying the virial product
with the virial factor. 

The Eddington luminosity of each object was calculated with the equation 
$L_{\rm Edd}=1.25 \times 10^{38} M_{\rm BH}$ (\ergs) \citep{wyi02}. 
We used the luminosity of the broad \Ha\ line as a proxy for the AGN bolometric luminosity 
Utilizing the relation between the luminosity of broad \Ha\ and the continuum luminosity at 5100\AA\ \citep{gre05}, and the bolometric correction 9.26 for L$_{5100}$ \citep{ric06}, 
we derived the bolometric luminosity as follows,
\begin{equation}\label{eq6}
L_{bol} =2.21 \times 10^{44} \left(\frac{L_{H\alpha}} {10^{42}\ {\rm erg\,s^{-1}}}\right)^{0.86} {\rm erg\,s^{-1}} ~ .
\end{equation}

\begin{figure}[t!]
\centering
\includegraphics[width=82mm]{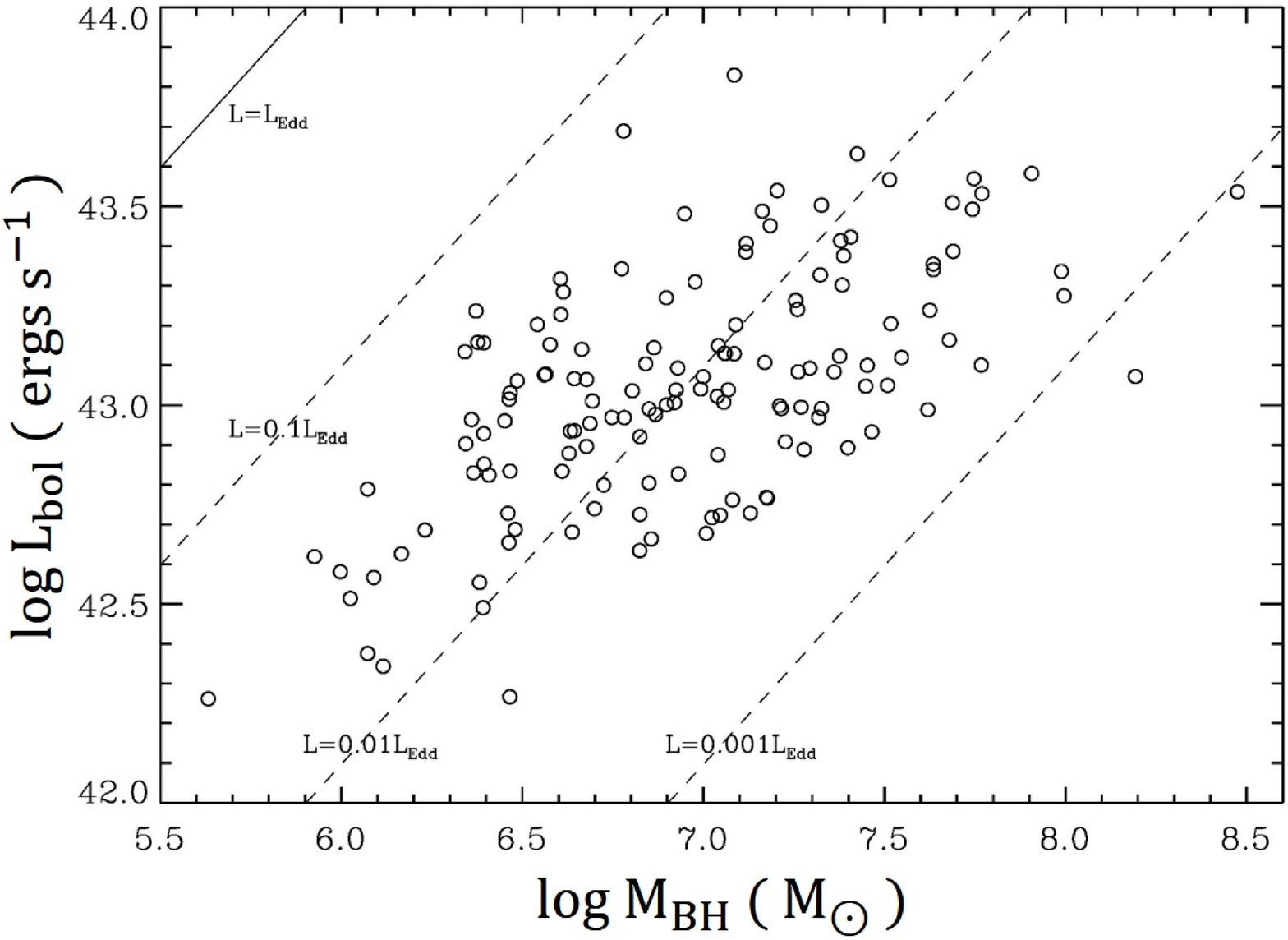}
\caption{Bolometric luminosity vs. black hole mass for 142 misclassified type 1 AGNs.
Dashed lines represent fixed Eddington ratios.
}
\label{fig4}
\end{figure}

\section{Results}

\subsection{Sample Properties}

Out of the parent sample of 4\,113 type 2 AGNs in the local universe ($0.02<z<0.05$), 
we find a total of 142 misclassified type 1 AGNs based on the presence of the broad
component of \Ha\ (FWHM $>$ 1000 \kms). 
The fraction of misclassified AGNs is $\sim$3.5\% of the parent 
sample. 25 objects have been previously referenced in the literature 
\citep[see][]{gre04,gre07,xia11,rei13}, indicating a novel discovery of missing type 1 AGNs. 

The distribution of the measured properties, i.e., the FWHM and luminosity of the broad \Ha\ component, and black hole mass, are presented in Figure \ref{fig3}.
The newly identified type 1 AGNs have a large range of the braod \Ha\ FWHM velocities,
ranging from 1700 to 19\,090 \kms. The mean line width of broad \Ha\ is log~FWHM$_{H\alpha} = 3.52 \pm 0.21$. In the case of the luminosity, all candidates are relatively low-luminosity AGNs
with \Ha\ luminosity lower than 10$^{42}$ \ergs. The mean broad \Ha\ luminosity is 
log L$_{H\alpha}=40.51\pm0.34$. 
The black hole mass of the sample ranges from $\sim$10$^{5.5}$ 
to $\sim$10$^{8.5}$ M$_{\odot}$ with a mean log M$_{\rm BH}$/M$_{\odot}$ = $6.94\pm0.51$, 
while the mean Eddington ratio is  log L$_{bol}$/L$_{Edd}$) = $-2.00\pm0.40$. 
Compared to the local supermassive black hole population, the misclassified AGNs
have a similar black hole mass range with a peak at $\sim$10$^{7}$ M$_{\odot}$ 
\citep{Heckman14}, while the Eddington ratio of the sample is relatively lower than
the one of high luminosity QSOs. 
Table \ref{tbl1} provides the list of 142 misclassified type 1 AGNs along with the measured
physical parameters. 
Figure \ref{fig4} presents the distribution of the misclassified type 1 AGNs in
the $L_{bol}-M_{BH}$ plane. Overall, the distribution of the sample is similar
to that of low luminosity type 1 AGNs \citep[e.g.,][]{wu02}.

\subsection{Comparison with MPA-JHU}

\begin{figure}[t!]
\centering
\includegraphics[width=70mm]{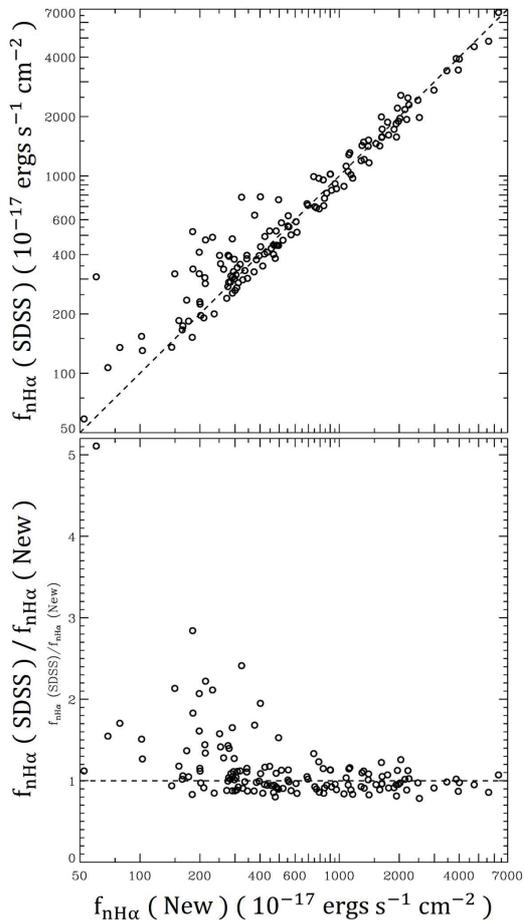}
\caption{Comparison of the narrow \Ha\ flux measurements of 142 misclassified type 1 AGNs 
adopted from the MPA-JHU catalog and from this study.}
\label{fig5}
\end{figure}

In this section, we compare our new measurements of the \Ha\ line luminosity with
that from the the MPA-JHU catalog as shown in Figure \ref{fig5}. 
Since the broad component of \Ha\ was included in our analysis, we expect that
our measurements of the narrow \Ha\ line luminosity is reliable, while the MPA-JHU
measurements are likely to be overestimated if the broad \Ha\ is not considered
in the fitting process.  

We find a clear trend that the narrow \Ha\ emission line flux measurements provided by 
the MPA-JHU catalog is larger than that of our measurement, particularly at the
low luminosity regime. The overall overestimation of the narrow \Ha\ line flux
can be interpreted as the contribution of the broad \Ha, since the narrow and broad
components were not decomposed, although the line flux can vary depending on
how the continuum around the \Ha\ region was determined. 
The difference is less significant for AGNs with a strong narrow \Ha\ line, 
presumably due to the weaker contribution of the broad \Ha\ component. 
These results demonstrate that the line flux measurements of the narrow \Ha\ 
can be significantly uncertain when a broad component is present
and that decomposing and subtracting a broad \Ha\ component from narrow lines (i.e., \Ha\ 
and \NII) is necessary for misclassified type 1 AGNs.      

\subsection{Velocity offset of the emission lines}

\begin{figure*}[t!]
\centering
\includegraphics[width=140mm]{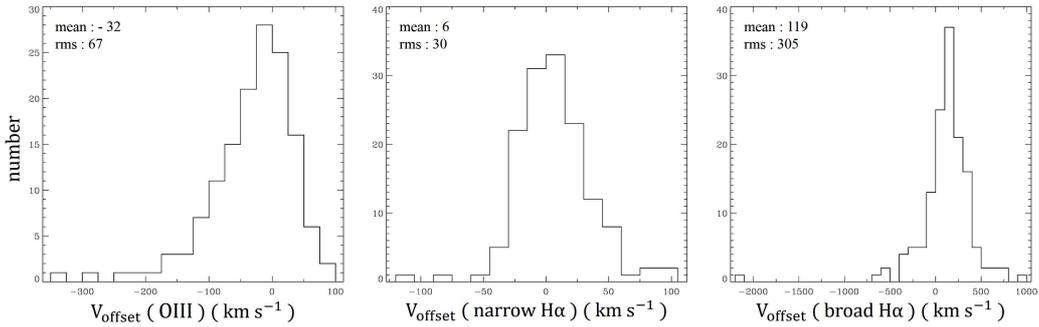}
\caption{Distribution of the velocity offset with respect to the systemic velocity
measured from stellar absorption lines. The \OIII\ line (left) presents on average
a larger velocity offset than the narrow component of \Ha\ (center). The broad component
of \Ha\ shows a significantly larger velocity offset than the narrow lines (right). 
In each panel, the mean and rms of the velocity offset is presented.
}
\label{fig6}
\end{figure*}

We investigate the kinematic properties of the ionized gas, 
by calculating the velocity offset of the [\OIII] and the \Ha\ lines
with respect to the systematic velocity, which is measured from stellar absorption lines. 
In the case of \Ha, we use the narrow and broad components separately, since 
the physical scale of these two components is clearly different. 
In Figure \ref{fig6}, we present the distribution of the velocity offset for each line. 
For the \OIII\ line, we used the flux centroid of the line profile as the velocity of the line
if a double Gaussian model was used for the fit. 
The \OIII\ line shows large velocity offsets ranging from $\sim-350$ to $\sim$100 \kms,
with a mean $-32$ \kms\ and rms 67 \kms. In addition, the distribution of the [\OIII] 
velocity offset is asymmetric, indicating that more than a half of the objects 
has blueshifted [\OIII] with relatively large velocities. 
The detected velocity offset with respect to the systemic velocity can be 
interpreted as due to outflows in the narrow-line region,
as various previous studies have used the \OIII\ velocity offset 
as an outflow indicator \citep[e.g.,][]{bo05,ko08,cr10, bw14}.

In the case of the narrow \Ha\ component, most galaxies show relatively small velocity offsets 
with a mean 6~\kms\ and rms 30 \kms. Given the uncertainty of the emission line
velocity, which is close to 10--20 \kms\ as measured from simulated mock spectra
based on the SDSS spectra by \citet{bw14}, 
only a small fraction of the sample seems to show a significant \Ha\ velocity offset.
These results of \OIII\ and \Ha\ velocity offsets are  
consistent with previous studies of type 2 AGNs \citep{ko08,bw14}.

In contrast, we detected large velocity offsets in the broad component of\Ha, 
with a mean velocity 119 \kms\ and with a rms of 305 \kms, 
which is much larger than that of narrow emission lines (\Ha\ and \OIII).
The nature of the velocity offset of the broad line is not clear without spatially resolved
measurements. We speculate that it may be due to the orbital motion of the black hole 
and accompanied BLR gas or the inflow/outflow motion of the gas in the BLR. 
More detailed studies are required to identify the nature of the velocity offset of broad emission lines.

\subsection{Kinematics of the ionized gas}

\begin{figure}[!t]
\centering
\includegraphics[width=70mm]{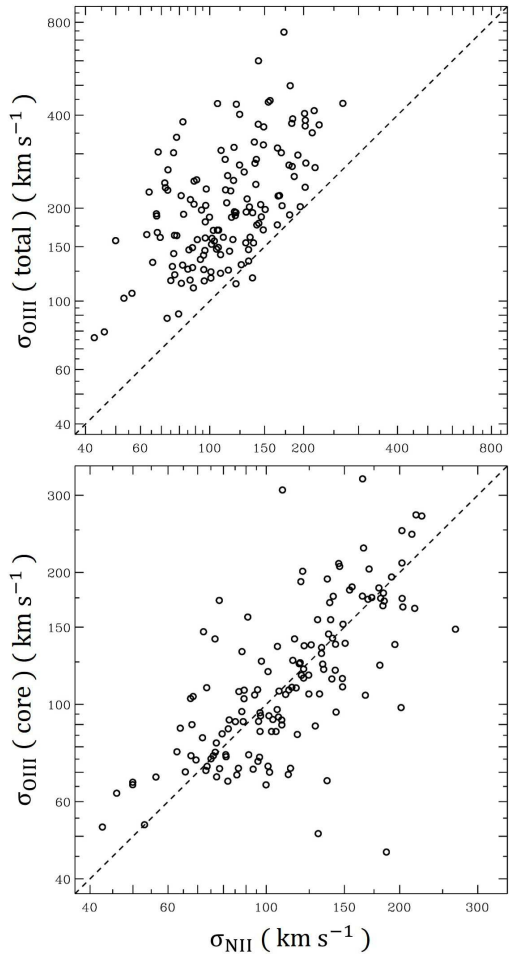}
\caption{Comparison of the line dispersion of \NII\ with that of total (top) 
and core component of [\OIII] (bottom). While the \OIII\ line is clearly broader
than \NII\ due to the strong wing component, the core component of \OIII\ shows
consistent line dispersion compared to \NII. 
}
\label{fig7}
\end{figure}

In Figure \ref{fig7} we present the velocity dispersion of the ionized gas.
For \OIII\ we calculated the second moment of the total line profile as
the velocity dispersion when we used a double Gaussian model for the fit.
In the case of \NII\ and \SII, we used the line dispersion of the best-fit Gaussian
model. Since we used the same Gaussian model for \NII\ and \SII, we only present
the velocity dispersion of \NII. 

Compared to the \NII\ lines, the velocity dispersion of \OIII\ is much larger, 
by a factor 1.8 in average. The larger \OIII\ velocity dispersion is expected from 
the presence of a wing component in the line profile. 
However, once we remove the wing component and measure
the velocity dispersion from the narrower core component from the best-fit double 
Gaussian models (bottom panel in Figure \ref{fig7}),
the velocity dispersions of \NII\ and \OIII\ become consistent albeit with significant 
scatter ($\sim$30\%). 
The mean difference of the velocity dispersion between \NII\ and the \OIII\ core component 
is only $\sim$1\%, confirming that the core component of \OIII\
and low-ionization lines (i.e., \NII\ and \SII) have similar kinematics,
presumably governed by the gravitational potential of the galaxy bulge 
\citep[see also][]{ko08}.

\begin{figure}[!t]
\centering
\includegraphics[width=70mm]{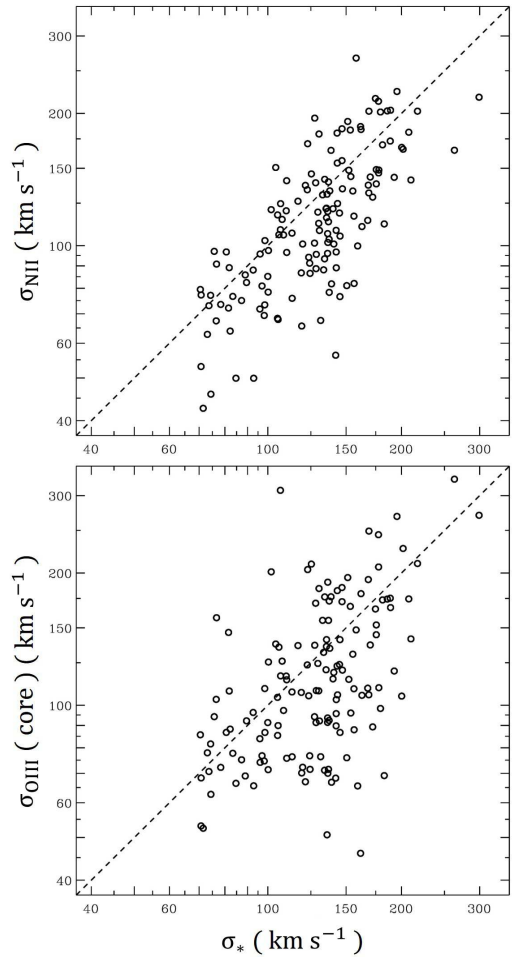}
\caption{Comparison of the stellar velocity dispersion with the velocity dispersions
of \NII\ (top) and the core component of [\OIII] (bottom). 
}
\label{fig8}
\end{figure}

In Figure \ref{fig8} we directly compare the velocity dispersion of \NII\ and the core
component of \OIII\ with stellar velocity dispersion. 
The velocity dispersion of \NII\ is on average smaller than the stellar velocity
dispersion by 14\%, while the scatter is $\sim$29\%.  
In the case of the core component of \OIII\ (bottom panel in Figure 8), 
the velocity dispersion is also smaller than the stellar velocity dispersion by 13\%, 
however the scatter is considerably larger ($\sim$43\%).
These results suggest that the velocity dispersions of narrow emission lines are 
on average consistent with stellar velocity dispersions, confirming the results
of previous studies \citep[e.g.,][]{nelson95}. Thus, the width of \NII\ (or 
\OIII\ core component) may be used as a proxy for stellar velocity dispersion.
However, if the velocity dispersions of the narrow-emission lines are used 
instead of the directly measured stellar velocity dispersions, significantly larger
uncertainty would be introduced due to the large scatter shown in Figure 8. 

The large scatter between emission and absorption lines can be attributed
to the contribution from outflow or inflow motion of the gas in the NLR,
which seems to preferentially affect the \OIII-emitting gas. 
In addition, we note that the measured stellar velocity dispersion from the 
SDSS spectra, which were extracted with a $3''$ aperture, can suffer 
from rotational broadening in the stellar absorption lines, depending on the 
orientation of the stellar disk and the relative strength of the rotation 
and random velocities as demonstrated by the spatially resolved measurements
\citep{kang13,woo13} and the simulated results with various line-of-sight 
measurement \citep{bell14}. 
The larger stellar velocity dispersion relative to the velocity dispersion of
\NII, by 0.06 dex (14\%), may indicate an overestimate of
the stellar velocity dispersion due to the rotation/inclination effect.



\begin{table*}
\centering
\caption{Physical parameters of 142 missing type 1 AGNs \label{tbl1}}
\begin{tabular}{lccccccr}
\toprule
AGN & $z$ & FWHM$_{H\alpha}$ & L$_{H\alpha}$ & L$_{bol}$ & $VP$ & L$_{bol}/L_{Edd}$ & $\sigma_{*}$ \\ 
(1) & (2) & (3) & (4) & (5) & (6) & (7) & (8) \\
\midrule
J004730.33+154149.4 & 0.0315 & 3460 & 40.1 & 42.7 & 6.05 & $-$2.20 & 134\\
J005847.49$-$010549.6 & 0.0465 & 5090 & 40.6 & 43.1 & 6.59 & $-$2.37 & 128\\
J010606.92+002025.1 & 0.0436 & 2880 & 40.1 & 42.7 & 5.87 & $-$2.05 & 73\\
J013402.54$-$094626.9 & 0.0408 & 4530 & 41.1 & 43.6 & 6.74 & $-$2.04 & 143\\
J015612.53+145423.9 & 0.0269 & 1930 & 40.5 & 43.0 & 5.70 & $-$1.53 & 98\\
J030834.31+003303.3 & 0.0308 & 3120 & 40.4 & 43.0 & 6.10 & $-$1.99 & 168\\
J031142.21+000853.3 & 0.0370 & 12990 & 40.5 & 43.1 & 7.42 & $-$3.22 & 128\\
J032525.35$-$060837.9 & 0.0345 & 2400 & 41.4 & 43.8 & 6.31 & $-$1.35 & 145\\
J033957.86$-$061215.1 & 0.0498 & 4520 & 40.5 & 43.0 & 6.44 & $-$2.31 & 136\\
J074507.25+460420.6 & 0.0313 & 6120 & 40.5 & 43.0 & 6.74 & $-$2.56 & 158\\
J075151.88+494851.5 & 0.0244 & 2740 & 40.4 & 43.0 & 5.98 & $-$1.87 & 113\\
J075217.05+254008.7 & 0.0454 & 2960 & 40.2 & 42.8 & 5.95 & $-$2.02 & 145\\
J075643.72+445124.1 & 0.0498 & 3540 & 40.6 & 43.1 & 6.29 & $-$2.03 & 108\\
J075828.11+374711.8 & 0.0408 & 2710 & 40.8 & 43.3 & 6.13 & $-$1.72 & 263\\
J080421.30+100610.9 & 0.0342 & 3530 & 40.6 & 43.1 & 6.29 & $-$2.02 & 134\\
J081726.41+101210.1 & 0.0457 & 5640 & 40.7 & 43.2 & 6.75 & $-$2.41 & 143\\
J082004.39+562320.6 & 0.0444 & 7930 & 40.6 & 43.1 & 7.00 & $-$2.76 & 113\\
J082351.90+421319.4 & 0.0376 & 1980 & 40.3 & 42.9 & 5.62 & $-$1.64 & 96\\
J082414.31+171955.0 & 0.0372 & 2010 & 40.7 & 43.2 & 5.84 & $-$1.48 & 110\\
J082620.81+055727.8 & 0.0448 & 3420 & 40.5 & 43.1 & 6.23 & $-$2.02 & 124\\
J084137.87+545506.5 & 0.0446 & 3200 & 41.0 & 43.5 & 6.39 & $-$1.77 & 162\\
J084143.50+013149.8 & 0.0499 & 3640 & 40.6 & 43.1 & 6.31 & $-$2.05 & 155\\
J090554.48+471045.5 & 0.0272 & 2550 & 40.3 & 42.8 & 5.84 & $-$1.87 & 183\\
J091330.33+565128.4 & 0.0410 & 2050 & 40.3 & 42.8 & 5.64 & $-$1.68 & 107\\
J091424.75+115625.5 & 0.0312 & 1350 & 40.0 & 42.6 & 5.16 & $-$1.40 & 74\\
J091708.26+292215.6 & 0.0353 & 3390 & 40.2 & 42.8 & 6.08 & $-$2.14 & 145\\
J092313.31+565622.2 & 0.0494 & 2810 & 40.5 & 43.0 & 6.03 & $-$1.86 & 172\\
J092528.24+233630.9 & 0.0330 & 4550 & 40.6 & 43.1 & 6.49 & $-$2.27 & 136\\
J093106.75+490447.1 & 0.0339 & 1980 & 40.4 & 43.0 & 5.68 & $-$1.59 & 147\\
J093551.60+612111.3 & 0.0393 & 2260 & 40.9 & 43.3 & 6.00 & $-$1.53 & 189\\
J093917.25+363343.8 & 0.0197 & 4020 & 40.3 & 42.9 & 6.27 & $-$2.26 & 151\\
J094319.15+361452.1 & 0.0221 & 2650 & 40.3 & 42.9 & 5.91 & $-$1.88 & 175\\
J094830.01+553822.6 & 0.0452 & 2100 & 40.5 & 43.1 & 5.79 & $-$1.58 & 100\\
J094931.37+343819.5 & 0.0388 & 2260 & 40.6 & 43.1 & 5.89 & $-$1.62 & 76\\
J095009.35+333409.5 & 0.0271 & 5000 & 40.2 & 42.8 & 6.41 & $-$2.51 & 137\\
J095437.22+063719.5 & 0.0410 & 6960 & 41.1 & 43.6 & 7.14 & $-$2.42 & 181\\
J095742.84+403315.9 & 0.0452 & 1820 & 40.3 & 42.9 & 5.57 & $-$1.54 & 105\\
J095824.97+103402.4 & 0.0417 & 4850 & 40.1 & 42.7 & 6.36 & $-$2.50 & 108\\
J100207.04+030327.6 & 0.0232 & 4580 & 40.8 & 43.3 & 6.61 & $-$2.18 & 119\\
J101439.55$-$004951.2 & 0.0491 & 3120 & 40.6 & 43.1 & 6.16 & $-$1.93 & 200\\
J101640.57+025125.2 & 0.0483 & 4370 & 40.1 & 42.7 & 6.24 & $-$2.43 & 120\\
J101833.47+141241.1 & 0.0323 & 3460 & 40.5 & 43.0 & 6.22 & $-$2.05 & 135\\
J101846.09+345001.6 & 0.0349 & 1650 & 40.6 & 43.2 & 5.62 & $-$1.33 & 77\\
J104243.85+314121.8 & 0.0345 & 3650 & 40.0 & 42.6 & 6.05 & $-$2.29 & 74\\
J104250.27+254616.3 & 0.0299 & 3360 & 41.0 & 43.5 & 6.41 & $-$1.83 & 105\\
J104546.94+371240.7 & 0.0241 & 3040 & 39.6 & 42.3 & 5.69 & $-$2.30 & 135\\
J104809.69+565459.4 & 0.0463 & 3790 & 40.5 & 43.0 & 6.29 & $-$2.15 & 124\\
J104930.92+225752.3 & 0.0327 & 2300 & 40.5 & 43.1 & 5.87 & $-$1.67 & 163\\
J105214.95+300328.3 & 0.0345 & 1940 & 40.5 & 43.1 & 5.71 & $-$1.52 & 136\\
J105427.88+330943.4 & 0.0433 & 1960 & 40.8 & 43.3 & 5.84 & $-$1.43 & 147\\
J105756.81+165434.4 & 0.0301 & 6250 & 40.4 & 42.9 & 6.69 & $-$2.63 & 119\\
J105833.33+461604.8 & 0.0399 & 3500 & 40.7 & 43.2 & 6.32 & $-$1.98 & 142\\
J110501.98+594103.5 & 0.0338 & 2530 & 41.0 & 43.5 & 6.18 & $-$1.56 & 137\\
J111106.73+073459.8 & 0.0479 & 4690 & 40.6 & 43.1 & 6.52 & $-$2.30 & 186\\
J111117.95+113315.8 & 0.0381 & 5880 & 40.9 & 43.4 & 6.86 & $-$2.38 & 162\\
J111258.29+215434.0 & 0.0296 & 5210 & 40.3 & 42.9 & 6.51 & $-$2.49 & 121\\
J111349.74+093510.7 & 0.0292 & 3940 & 41.2 & 43.6 & 6.65 & $-$1.89 & 210\\
J111653.41+275822.8 & 0.0347 & 3680 & 40.5 & 43.0 & 6.27 & $-$2.11 & 104\\
J111926.22+031205.7 & 0.0236 & 1960 & 39.7 & 42.3 & 5.34 & $-$1.87 & 82\\
J112008.68+341845.8 & 0.0368 & 2540 & 40.3 & 42.9 & 5.86 & $-$1.85 & 102\\
\bottomrule
\end{tabular}
\end{table*}


\setcounter{table}{0}
\begin{table*}
\centering
\caption{\emph{continued}}
\begin{tabular}{lccccccr}
\toprule
AGN & $z$ & FWHM$_{H\alpha}$ & L$_{H\alpha}$ & L$_{bol}$ & $VP$ & L$_{bol}/L_{Edd}$ & $\sigma_{*}$ \\ 
(1) & (2) & (3) & (4) & (5) & (6) & (7) & (8) \\
\midrule
J112011.14+340858.9 & 0.0352 & 2410 & 40.1 & 42.7 & 5.71 & $-$1.89 & 71\\
J112135.17+042647.2 & 0.0470 & 1780 & 40.4 & 43.0 & 5.59 & $-$1.49 & 100\\
J112637.73+513423.0 & 0.0265 & 1760 & 40.0 & 42.6 & 5.39 & $-$1.64 & 93\\
J112726.64+264051.5 & 0.0329 & 4860 & 40.4 & 42.9 & 6.46 & $-$2.42 & 129\\
J113355.93+670107.0 & 0.0397 & 6130 & 40.9 & 43.4 & 6.92 & $-$2.40 & 156\\
J113409.01+491516.3 & 0.0373 & 2410 & 40.1 & 42.7 & 5.69 & $-$1.91 & 70\\
J113543.71+490959.8 & 0.0353 & 2450 & 39.9 & 42.5 & 5.62 & $-$2.00 & 105\\
J114223.58+153340.9 & 0.0444 & 1940 & 40.5 & 43.0 & 5.69 & $-$1.55 & 82\\
J114530.25+094344.7 & 0.0214 & 3080 & 40.4 & 42.9 & 6.05 & $-$2.00 & 159\\
J114612.17+202329.9 & 0.0233 & 5190 & 40.4 & 43.0 & 6.55 & $-$2.45 & 195\\
J114840.42$-$001710.3 & 0.0474 & 3210 & 40.5 & 43.0 & 6.15 & $-$1.98 & 208\\
J115246.29+232833.6 & 0.0220 & 5580 & 40.6 & 43.1 & 6.68 & $-$2.45 & 169\\
J115623.23+195932.1 & 0.0409 & 3040 & 40.4 & 43.0 & 6.08 & $-$1.96 & 136\\
J120144.91+201941.9 & 0.0234 & 1940 & 40.3 & 42.8 & 5.59 & $-$1.63 & 123\\
J120234.39+544500.7 & 0.0499 & 4050 & 40.6 & 43.1 & 6.40 & $-$2.16 & 93\\
J120443.31+311038.2 & 0.0249 & 4270 & 40.9 & 43.4 & 6.61 & $-$2.06 & 127\\
J120704.75+090647.9 & 0.0344 & 6150 & 41.1 & 43.5 & 7.00 & $-$2.33 & 168\\
J120908.80+440011.4 & 0.0376 & 2030 & 40.6 & 43.2 & 5.81 & $-$1.52 & 143\\
J124054.96+080323.1 & 0.0477 & 1620 & 40.6 & 43.2 & 5.61 & $-$1.32 & 97\\
J124240.45+092851.8 & 0.0240 & 2330 & 39.9 & 42.6 & 5.61 & $-$1.92 & 76\\
J124610.76+275615.9 & 0.0231 & 4840 & 40.5 & 43.0 & 6.50 & $-$2.37 & 123\\
J124954.76+503453.9 & 0.0469 & 2490 & 40.4 & 42.9 & 5.87 & $-$1.80 & 133\\
J125258.13+090157.2 & 0.0400 & 3770 & 40.5 & 43.0 & 6.30 & $-$2.13 & 175\\
J125552.36+103055.2 & 0.0484 & 4400 & 40.9 & 43.4 & 6.62 & $-$2.11 & 130\\
J130340.81+534323.6 & 0.0276 & 1670 & 40.0 & 42.6 & 5.32 & $-$1.62 & 136\\
J130342.82+324824.5 & 0.0367 & 1820 & 40.1 & 42.7 & 5.46 & $-$1.64 & 74\\
J130422.19+361543.1 & 0.0443 & 1850 & 41.3 & 43.7 & 6.01 & $-$1.19 & 137\\
J130620.97+531823.1 & 0.0237 & 8830 & 40.8 & 43.3 & 7.22 & $-$2.75 & 141\\
J130705.01+024337.1 & 0.0477 & 4570 & 40.4 & 43.0 & 6.44 & $-$2.32 & 147\\
J130737.68+433117.7 & 0.0353 & 3420 & 40.6 & 43.1 & 6.27 & $-$1.99 & 83\\
J132209.96+331005.6 & 0.0373 & 1490 & 40.0 & 42.6 & 5.23 & $-$1.51 & 78\\
J132336.84+062424.3 & 0.0394 & 2590 & 40.4 & 43.0 & 5.92 & $-$1.83 & 85\\
J133204.81+170256.2 & 0.0215 & 4140 & 40.7 & 43.2 & 6.49 & $-$2.12 & 152\\
J134244.42+350346.4 & 0.0243 & 2300 & 40.1 & 42.7 & 5.69 & $-$1.83 & 87\\
J134249.94+294546.4 & 0.0431 & 2900 & 40.8 & 43.3 & 6.21 & $-$1.76 & 170\\
J135419.95+325547.7 & 0.0261 & 4220 & 40.8 & 43.3 & 6.55 & $-$2.09 & 177\\
J135747.63+072346.5 & 0.0236 & 3720 & 40.1 & 42.7 & 6.09 & $-$2.29 & 100\\
J140543.16+251352.9 & 0.0298 & 6230 & 40.7 & 43.2 & 6.85 & $-$2.48 & 139\\
J140718.28+125313.9 & 0.0274 & 4980 & 40.2 & 42.8 & 6.40 & $-$2.50 & 139\\
J140804.00+071939.4 & 0.0239 & 1890 & 40.4 & 42.9 & 5.62 & $-$1.56 & 82\\
J141057.23+252950.0 & 0.0310 & 2400 & 40.5 & 43.1 & 5.91 & $-$1.71 & 175\\
J141809.22+073352.3 & 0.0247 & 4500 & 40.2 & 42.8 & 6.31 & $-$2.42 & 299\\
J142042.91+262503.2 & 0.0379 & 1900 & 40.8 & 43.3 & 5.83 & $-$1.38 & 130\\
J142255.33+325102.3 & 0.0342 & 4060 & 40.8 & 43.3 & 6.48 & $-$2.09 & 140\\
J142307.51+283542.3 & 0.0293 & 2810 & 40.6 & 43.1 & 6.07 & $-$1.83 & 130\\
J142704.54+355409.5 & 0.0290 & 5930 & 40.9 & 43.3 & 6.86 & $-$2.39 & 179\\
J143318.47+344404.4 & 0.0343 & 3180 & 40.5 & 43.0 & 6.13 & $-$1.99 & 202\\
J143727.85+254556.0 & 0.0328 & 1200 & 39.6 & 42.3 & 4.86 & $-$1.47 & 136\\
J144331.28+492335.1 & 0.0302 & 6920 & 40.6 & 43.2 & 6.91 & $-$2.61 & 137\\
J144629.99+500130.7 & 0.0427 & 6110 & 41.0 & 43.5 & 6.97 & $-$2.35 & 132\\
J144837.03+514331.0 & 0.0364 & 2170 & 40.3 & 42.8 & 5.69 & $-$1.73 & 71\\
J145048.68+200301.4 & 0.0437 & 5170 & 40.4 & 43.0 & 6.56 & $-$2.43 & 154\\
J150214.29+182928.7 & 0.0477 & 2090 & 40.5 & 43.1 & 5.79 & $-$1.58 & 81\\
J150511.42$-$020831.0 & 0.0372 & 5950 & 40.3 & 42.9 & 6.63 & $-$2.60 & 138\\
J150653.38+125131.2 & 0.0216 & 1840 & 39.7 & 42.4 & 5.30 & $-$1.79 & 106\\
J150656.41+125048.6 & 0.0224 & 2980 & 40.2 & 42.7 & 5.93 & $-$2.06 & 217\\
J151512.25+152412.3 & 0.0457 & 1540 & 40.7 & 43.2 & 5.60 & $-$1.23 & 98\\
J151907.55+260750.6 & 0.0447 & 2520 & 40.5 & 43.0 & 5.92 & $-$1.78 & 117\\
J154357.33+283126.4 & 0.0323 & 1580 & 40.6 & 43.1 & 5.57 & $-$1.30 & 110\\
J154744.14+412408.2 & 0.0326 & 1900 & 40.7 & 43.2 & 5.77 & $-$1.44 & 89\\
\bottomrule
\end{tabular}
\end{table*}


\setcounter{table}{0}
\begin{table*}
\centering
\caption{\emph{continued}}
\begin{tabular}{lccccccr}
\toprule
AGN & $z$ & FWHM$_{H\alpha}$ & L$_{H\alpha}$ & L$_{bol}$ & $VP$ & L$_{bol}/L_{Edd}$ & $\sigma_{*}$ \\ 
(1) & (2) & (3) & (4) & (5) & (6) & (7) & (8) \\
\midrule
J155926.11+521235.3 & 0.0423 & 4380 & 40.9 & 43.4 & 6.64 & $-$2.08 & 134\\
J160154.01+315331.3 & 0.0450 & 3240 & 40.9 & 43.4 & 6.35 & $-$1.83 & 178\\
J160417.30+042135.7 & 0.0465 & 3660 & 40.3 & 42.8 & 6.16 & $-$2.20 & 124\\
J160445.00+444316.9 & 0.0433 & 2460 & 40.4 & 42.9 & 5.86 & $-$1.79 & 107\\
J160505.15+452634.8 & 0.0434 & 5870 & 41.1 & 43.6 & 6.98 & $-$2.27 & 178\\
J160652.16+275539.1 & 0.0463 & 3250 & 41.1 & 43.5 & 6.43 & $-$1.76 & 128\\
J160746.71+253214.9 & 0.0406 & 1440 & 40.2 & 42.8 & 5.30 & $-$1.38 & 96\\
J161243.05+002247.9 & 0.0445 & 9230 & 40.8 & 43.3 & 7.22 & $-$2.82 & 189\\
J161630.67+354228.9 & 0.0278 & 5730 & 40.5 & 43.0 & 6.68 & $-$2.50 & 151\\
J162131.63+245952.4 & 0.0378 & 3250 & 40.5 & 43.0 & 6.15 & $-$2.01 & 110\\
J162938.37+384139.2 & 0.0356 & 1600 & 39.9 & 42.5 & 5.25 & $-$1.61 & 71\\
J163453.66+231242.6 & 0.0387 & 3810 & 41.0 & 43.5 & 6.55 & $-$1.92 & 144\\
J164107.63+224924.8 & 0.0339 & 2810 & 40.6 & 43.1 & 6.09 & $-$1.82 & 89\\
J164520.61+424528.0 & 0.0494 & 5050 & 40.6 & 43.1 & 6.60 & $-$2.35 & 98\\
J171518.57+573931.6 & 0.0281 & 4340 & 40.1 & 42.7 & 6.25 & $-$2.40 & 131\\
J173159.21+595817.9 & 0.0292 & 4440 & 40.1 & 42.7 & 6.28 & $-$2.42 & 153\\
J173809.30+584253.6 & 0.0288 & 6130 & 40.6 & 43.1 & 6.78 & $-$2.52 & 125\\
J204745.25$-$052515.6 & 0.0458 & 3200 & 40.9 & 43.4 & 6.35 & $-$1.81 & 156\\
J210917.64+001619.1 & 0.0500 & 13500 & 41.1 & 43.5 & 7.70 & $-$3.04 & 169\\
J230920.26+004523.3 & 0.0323 & 5690 & 41.0 & 43.5 & 6.92 & $-$2.28 & 142\\
J232337.44+133908.1 & 0.0425 & 2850 & 40.4 & 43.0 & 6.01 & $-$1.91 & 193\\
J232611.29+140148.1 & 0.0462 & 7190 & 40.4 & 43.0 & 6.85 & $-$2.73 & 143\\
\bottomrule
\end{tabular}
\tabnote{{\sc Notes.} Column 1: Object name. Column 2: Redshift. Column 3: FWHM velocity of
the broad component of the \Ha\ emission line in \kms. Column 4: Luminosity of 
the broad component of the \Ha\ emission line in log scale (\ergs). 
Column 5: Bolometric luminosity in log scale (\ergs). Column 6: Virial Product 
in log scale (M$_{\odot}$). Column 7: Eddington ratio in log scale. 
Column 8: Stellar velocity dispersion in \kms\ adopted from the SDSS catalogue.}
\end{table*}


\section{Discussion and Conclusions}

We presented a sample of 142 misclassified type 1 AGNs identified out of a parent sample of 4\,113 
type 2 AGNs at $0.02<z<0.05$, selected from the SDSS DR 7 based on the emission-line flux ratios. The fraction of the misclassified type 1 AGNs among type 2 AGNs is $\sim$3.5\%;
this should be considered as a lower limit because we conservatively identified the 
broad \Ha\ component by excluding AGNs that require a broad component with relatively small width. 
If we extrapolate the fraction of the misclassified type 1 AGNs, 3.5\%, to higher redshift, 
it is expected that a large number of AGNs could be misclassified as type 2 AGNs in the 
SDSS DR7 catalog. The discovery of the missing type 1 AGN population will shed light on
understanding the low-mass, low-luminosity AGNs and studying AGN population and 
their properties. 

The newly identified type 1 AGN sample has a mean black hole mass log (M$_{BH}$/M$_{\odot}$)=$6.94\pm0.51$ based on the most recent single-epoch black hole mass calibrator -- although 
the precise black hole mass values depend on the virial factor. The mean broad \Ha\ luminosity of the sample is log (L$_{H\alpha}$/(\ergs) = $40.50\pm0.35$, and the mean Eddington ratio is log L$_{bol}/L_{Edd}$ = $-2.00\pm0.40$. The low Eddington ratios of the sample
imply that the AGN continuum is too weak to change the host galaxy color, which is
consistent with the relatively red color of the targets shown in the SDSS images.

We investigated the velocity offset of each emission line with respect to the systemic velocity measured from stellar absorption lines. We find that the \OIII\ lines show relatively
large velocity offsets while the velocity offsets of the \Ha\ lines are weaker than for \OIII,
indicating that the \OIII-emitting gas is influenced by outflows more strongly.
The velocity dispersions of the \NII\ lines and the \OIII\ core components are
consistent with each other and with stellar velocity dispersions, albeit with significant scatter,
suggesting that the kinematics of the gas in the NLR
is mainly governed by the large scale gravitational potential (i.e., the galaxy bulge),
while outflow/inflow motions can vary the velocity dispersion
of narrow emission lines. The kinematic properties of the ionized gas of the sample
are similar to those of the general type 1 AGN population, supporting the idea that
these objects are simply misclassified type 1 AGNs.


\acknowledgments

This work was supported by the National Research Foundation of Korea (NRF) grant funded by the Korean government (MEST; No. 2012-R1A2A2A01006087). 
J.H.W. acknowledges the support by the Korea Astronomy and Space Science Institute (KASI) grant funded by the Korea government (MEST). 
by the Korea government (No. 2012-006087). 
J.G.K. acknowledges the support by the 2012 R\&E Program of Gyeonggi Science High School for the Gifted (GSHS).




\end{document}